\documentclass[11pt]{book} 
\usepackage{amsmath} % AMS Math Package
\usepackage{amsthm} % Theorem Formatting
\usepackage{amssymb} % Math symbols such as \mathbb
\usepackage{graphicx} % Allows for eps images
\usepackage{multicol} % Allows for multiple columns
\usepackage[dvips,letterpaper,margin=1in,bottom=1in]{geometry}
\usepackage{hyperref}
\usepackage{mathrsfs}
\usepackage{amsmath,amscd}
\usepackage[all,cmtip]{xy}
\usepackage{bbm}
\usepackage{lipsum}

\newcommand{\cC}{\ensuremath{\mathcal C}}
\newcommand{\cV}{\ensuremath{\mathcal V}}

\newcommand{\cP}{\ensuremath{\mathcal P}}

\newcommand{\bS}{\ensuremath{\mathbf S}}

\newcommand{\conc}{{\ensuremath{\|}}}

\newcommand*{\titleGM}{\begingroup % Create the command for including the title page in the document
\hbox{ % Horizontal box
\hspace*{0.2\textwidth} % Whitespace to the left of the title page
\rule{1pt}{\textheight} % Vertical line
\hspace*{0.05\textwidth} % Whitespace between the vertical line and title page text
\parbox[b]{0.75\textwidth}{ % Paragraph box which restricts text to less than the width of the page

{\noindent\huge\bfseries Modeling Software Development Methodologies \\}\\[2\baselineskip] % Title
{\large \textit{A Logic Based Approach}}\\[4\baselineskip] % Tagline or further description
{\large \textsc{ Farzad Mahdikhani \\
Mohammad Reza Abbasifard}} % Author name

\vspace{0.5\textheight} % Whitespace between the title block and the publisher
{\noindent University of Tehran \\
           FMRG.FM-MRA~160701 }\\[\baselineskip] % Publisher and logo
}}
\endgroup}

\makeatletter % Need for anything that contains an @ command 
\renewcommand{\maketitle} % Redefine maketitle to conserve space
{ \begingroup \vskip 10pt \begin{center} \Huge {\bf \@title}
    \vskip 10pt \large \@author \hskip 20pt \@date \end{center}
  \vskip 10pt \endgroup \setcounter{footnote}{0} }
\makeatother % End of region containing @ commands
\let\baraccent=\= % rename builtin command \= to \baraccent
\renewcommand{\=}[1]{\stackrel{#1}{=}} % for putting numbers above =

\newcount\colveccount
\newcommand*\colvec[1]{
        \global\colveccount#1
        \begin{pmatrix}
        \colvecnext
}
\def\colvecnext#1{
        #1
        \global\advance\colveccount-1
        \ifnum\colveccount>0
                \\
                \expandafter\colvecnext
        \else
                \end{pmatrix}
        \fi
}

\usepackage{cancel}

\theoremstyle{definition}

\theoremstyle{remark}

\newcommand\abstractname{Abstract}  %%% here
\makeatletter

  \newenvironment{abstract}{%
      \textwidth=150pt
      \null \vfil
      \@beginparpenalty\@lowpenalty
      \begin{center}%
        \bfseries \abstractname 
        \@endparpenalty 
        \@M
      \end{center}
      \itshape\leftskip=1.5cm
      
      \rightskip=1cm}%
     { \par \vfil\null}

\begin{document}
\titleGM

\begin{abstract}
In the last two decades, the growing trend of software development industry has made different aspects of software engineering more interesting for the computer science research community. Software development life-cycle is one of these aspects that has a significant impact on the success/failure of software development projects. Since each software development methodology relatively provides its own software development life-cycle, encoding software development life-cycle in a workflow representation language can help developers to properly manage their software projects. In addition, such encoding can also be used in CASE tools. In this report, we consider the software development life-cycle as a workflow that can be represented by semantic web and rule based languages. Such consideration let one analyze the properties of the life-cycle. Specifically, we take a well-known agent oriented software development methodology and show that its corresponding life-cycle can be specified by Transaction Logic easily and concisely.  Finally, the compact and clear representation of the life-cycle in Transaction Logic can be used in CASE tools to guide software developers.

\end{abstract}

\chapter{Introduction} \label{chap:intro}

Due to the significant impact of a software development process and workflow on its resulted software system, the development and consideration of these workflows always have been in the center of attention of software engineering researchers. A \emph{software development process}  is a collection of coordinated activities designed to construct a complex software system. Every software development pattern, e.g. agent oriented or object oriented, is supposed to provide a precise definition of its corresponding software development process. 

Since each methodology offers different modeling concepts, analysis and design techniques, notations and supporting tools, selection of the appropriate methodology is crucial for developers. In general, the vast range of available methodologies may cause undesirable effects: (1) Confusion in selecting a methodology for a specific problem, in particular for industrial developers (2) Difficulty in reaching standards due to multiplicity of methodologies using different specifications; (3) Lack of sufficient samples and case studies in existing methodologies. To address these problems some frameworks are needed to evaluate agent oriented methodologies from different views. As one of the main components of software development methodologies, the evaluation of development life-cycle is playing an important role in the evaluation of software development methodologies. The encoding of software development life-cycle in a workflow representation language can also be used for the evaluation of such life-cycles. Moreover, although several approaches \cite{BassedaCSICC2007,BassedaICEE2007,DBLP:conf/tase/AlinaghiGSB09,DBLP:conf/sysose/BassedaAG09,DBLP:conf/aiccsa/BassedaTAGM09} has been proposed to evaluate agent oriented methodologies based on the dependency of stages and artifacts, none of these approaches are used in agent-based CASE tools.

There are several studies for the comparison of software development life-cycles. A set of criteria for feature based analysis of methodologies has been provided by \cite{Sturm2004,DBLP:journals/ijseke/SturmTG08}. This set of criteria consists of software Engineering criteria and agent based characteristics. A similar work has been done in \cite{Dam2004}. In all of these studies, only a feature based comparison of methodologies have been provided and the life-cycle is not viewed as a workflow. In practice, although feature based studies can guide developers to choose an appropriate methodology for their software development projects, they cannot guide developers during the software development process. 

The recent developments in Semantic Web services and the successful projects such as the WSMO,\footnote{\textit{http://www.wsmo.org/}} OWL-S,\footnote{\textit{http://www.daml.org/services/owl-s/}} and SWSL\footnote{\textit{http://www.w3.org/Submission/SWSF-SWSL/}} have directed us to the logic based modeling of the software development life-cycles. Moreover, the recent developments in \cite{DBLP:conf/esws/RomanKF08,DBLP:conf/semweb/RomanK08,DBLP:conf/vldb/RomanK07} show that Concurrent Transaction Logic can be considered as a very powerful tool for representation of semantic web workflows. In this report, we will also show that the Transaction Logic also can be used for modeling software development life-cycles. Such modeling can be used in several software engineering aspects, e.g CASE tools or methodology evaluation. For example, there exists an interpreter \cite{Fodor:2010:TTL:1836089.1836115} for the execution of Concurrent Transaction Logic that can be used in the development of CASE tools.

The rest of this report is structured as follows: to make the report self-contained, in Chapter~\ref{chap:tr}, we provide a brief explanation of concurrent transaction logic; in Chapter~\ref{chap:methodology}, we will briefly review our methodlogy; Chapter~\ref{chap:case_study} provides an example as our case study; Finally Chapter~\ref{chap:conclusion} draws our conclusion. 

\chapter{Overview of Concurrent Transaction Logic} \label{chap:tr}

In this chapter we briefly remind the reader a number of standard concepts in logic and Transaction Logic. For further details, the reader is referred to \cite{Bonner95transactionlogic,trans-chapter-98,Bonner94anoverview,trans-iclp93,concurrent-tr-96}. 
%This material is borrowed from  \cite{DBLP:conf/lpnmr/Basseda15,DBLP:conf/padl/BassedaK15,DBLP:conf/rr/BassedaK15,DBLP:conf/rr/BassedaKB14} and is not our contribution. 

To represent workflow tasks and actions, we only assume denumerable sets of variables \cV, constants \cC, and predicate symbols $\cP$---all three sets being pairwise disjoint. Actions in Transaction Logic update the state of a system by adding or deleting statements about predicates. \textbf{\textit{Atomic formulas}} (or just \textbf{\emph{atoms}}) have the form $p(t_1,...,t_n)$, were $p \in \cP$ and each $t_i$ is either a constant or a variable. 

Transaction Logic is a faithful extension of the first-order predicate calculus and so all of that syntax carries over. In this report, since we need just rules, we explain related connectives and omit the rest of Transaction Logic's syntax from now on. In our application, one the useful connectives from Transaction Logic is the \textbf{\emph{serial conjunction}}, denoted  
$\otimes$. It is a binary associative connective, like the classical conjunction, but it is not commutative. Informally, the formula $\phi\otimes\psi$ is understood as a composite
action that denotes an \emph{execution} of $\phi$ followed by an execution of $\psi$. The \textbf{\emph{concurrent conjunction}} connective, $\phi\conc\psi$, is 
associative \emph{and commutative}. Informally, it says that $\phi$ and $\psi$ can execute in an \emph{interleaved} fashion. The logic also has other connectives
but they are beyond the scope of this report.

As pointed earlier, Transaction Logic provides a general, extensible mechanism of \textbf{\emph{elementary updates}} or elementary \textbf{\emph{actions},} which have the important effect of taking the infamous \emph{frame problem} out of many considerations in this logic (see \cite{trans-chapter-98,Bonner94anoverview,Bonner95transactionlogic}).
Here we will use only the following two types of elementary actions: $+p(t_1,\dots,t_n)$ and $-p(t_1,\dots,t_n)$, where $p$ denotes a predicate symbol of appropriate arity and $t_1$, ..., $t_n$ are terms. These actions can be formally defined as follows: Given a state $\bS$ and a \emph{ground} elementary action $\alpha = +p(a_1,\dots,a_n)$, an execution of $\alpha$ at state $\bS$ adds the literal $p(a_1,\dots,a_n)$. Similarly, executing $-p(a_1,\dots,a_n)$ results in a state that is exactly like $\bS$, but $p(a_1,\dots, a_n)$ is deleted.

A \textbf{\emph{serial rule}}   is a statement of the form
\begin{equation}\label{eq:serial-horn}
\begin{array}{l}
  h \leftarrow b_1 \otimes b_2 \otimes \ldots \otimes b_n .
\end{array}
\end{equation}
where $h$ is an atomic formula and $b_1$, ..., $b_n$ are literals or elementary actions. The informal meaning of such a rule is that $h$ is a complex action and one way to execute $h$ is to execute $b_1$ then $b_2$, etc., and finally to execute $b_n$.

\chapter{Representing Lifecycles in Transaction Logic} \label{chap:methodology}

A software development life-cycle is consisting of a set of tasks that are partially ordered due to their required and produced artifacts. In fact, each task either need to use some artifacts provided by other tasks, or is a starting task, i.e. a task can be started at the beginning of the project. 

To represent elements of a software engineering project, we assume denumerable sets of variables \cV, constants \cC, and predicate symbols $\cP$---all three sets being pairwise disjoint. In our case, constants are representing \emph{\textbf{components}} or \emph{\textbf{parts}} of the software or models. For instance, the agent $ seller \in \cC $ represents an agent selling stuff in a virtual multi-agent marketplace. 

Like other formal logical representations, \textbf{\textit{atomic formulas}} (or just \textbf{\emph{atoms}}) have the form $p(t_1,...,t_n)$, were $p \in \cP$ and each $t_i$ is either a constant
or a variable. A \textbf{\emph{fact}} is a \textbf{\emph{ground}} (i.e., variable-free) atom. Every software engineering task $ \theta $ can be modeled by the following components:

\begin{itemize}
\item $p_\theta(X_1,...,X_n)$ is an atom in which $X_1,...,X_n$ are variables and $p_\theta$ is a predicate that is reserved to represent the software development task and can be used for no other purpose;
\item $Req_{\theta}$, called the \textbf{requirements} of $\theta$, is a set of literals that are representing artifacts required for the completion of this task;
\item $Prod_{\theta}$, called the {\textbf{productions}} of $\theta$, is a set of literals that are representing artifacts/requirements produced by the completion of this task or requirements that is deleted by its completion;
\end{itemize}

This model of actions and artifacts can represent different types of dependencies between software development tasks. Finally, a software development life-cycle can be represented as a composite action in Transaction Logic, defining the sequence of different tasks. Tasks are also ordered based on their corresponding required/produced artifacts. 

This technique can help us to extract the life-cycle of a software development methodology based on the dependency between artifacts. Conjunction and disjunction are the two types of
dependencies between artifacts which are described as follows:

\section{Conjunction Dependency}

\begin{figure}
  \centering
  \includegraphics[scale=.35]{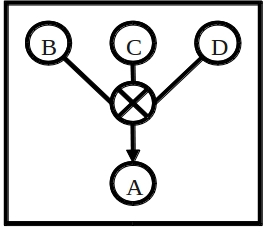}
  \caption{Conjunction Dependency.}
  \label{fig:Conjunction}
\end{figure}

Sometimes, in software engineering methodologies' lifecycle, completeness of some other models are required for preparing a model. For example, in Figure~\ref{fig:Conjunction}, model A has conjunction dependency on model B, C and D. Therefore Model A can not be started till B, C and D are provided. In this type of dependency, Complexity of the dependant is influenced by complexity of every single one of its prerequisites. Actions of this type can be represented as follows:

\begin{equation}
\alpha \longleftarrow (B \wedge C \wedge D) \otimes + A.
\end{equation}

\section{Disjunction Dependency}

\begin{figure}
  \centering
  \includegraphics[scale=.35]{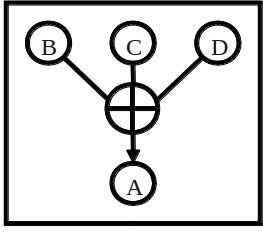}
  \caption{Disjunction Dependency.}
  \label{fig:Disjunction}
\end{figure}

In the agent oriented methodology lifecycle, sometimes there is another type of dependencies between models, in which completeness of one of the other models is adequate for preparing the particular model. For example, in Figure~\ref{fig:Disjunction}, model A has disjunction dependency on model B, C and D. In that case model A can be started as soon as one of B, C or D is provided. Actions of this type can be represented as follows:

\begin{align}
\alpha \longleftarrow B  \otimes + A. \\ \nonumber
\alpha \longleftarrow C  \otimes + A. \\ \nonumber
\alpha \longleftarrow D  \otimes + A. 
\end{align}

\section{Monitoring the Workflow}

As mentioned in Chapter~\ref{chap:intro}, the provided models can be applied for several purposes, including management of software development process. To monitor the progress of a software development project, project managers need to check the status of different artifacts and components and figure out critical dependencies in the project life-cycle. Depending on the size of projects, sometimes managers face a large number of tasks and artifacts that can be difficult to manage and monitor. Our proposed model can be used simply to help managers to control the dynamics of project, while track the status of artifacts and possibility of starting a task. 

\begin{figure}
  \centering
  \includegraphics[scale=.35]{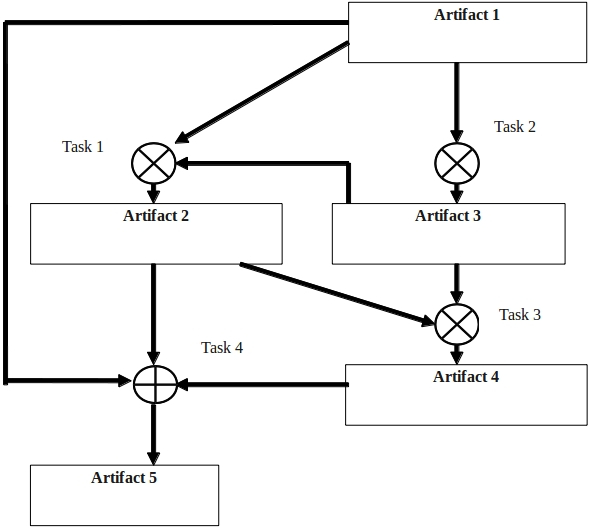}
  \caption{A sample task network from a software development methodology.}
  \label{fig:Sample_01}
\end{figure}

Let us explain this application via a simple example. Consider the simple task network shown in Figure~\ref{fig:Sample_01}. This task network can be represented using the Transaction Logic rules in~(\ref{eq:sample-rule-01}),~(\ref{eq:sample-rule-02}),~(\ref{eq:sample-rule-03}), and~(\ref{eq:sample-rule-04}). Let $ C $ in these rules denote the under development component of the projects. 

\begin{align} \label{eq:sample-rule-01}
Task_1(C) \longleftarrow &~(~Artifact_1(C) \wedge \\ \nonumber 
                      &~~~Artifact_3(C)~) \\ \nonumber 
                      &\otimes ~~~+ Artifact_2(C).
\end{align}

\begin{align} \label{eq:sample-rule-02}
Task_2(C) \longleftarrow &~(~Artifact_1(C)~) \\ \nonumber 
                      &\otimes ~~~+ Artifact_3(C).
\end{align}

\begin{align} \label{eq:sample-rule-03}
Task_3(C) \longleftarrow &~(~Artifact_2(C) \wedge \\ \nonumber 
                      &~~~Artifact_3(C)~) \\ \nonumber 
                      &\otimes ~~~+ Artifact_4(C).
\end{align}

\begin{align} \label{eq:sample-rule-04}
Task_4(C) \longleftarrow &~(~Artifact_1(C)~) \\ \nonumber 
                      &\otimes ~~~+ Artifact_5(C).  \\ \nonumber 
Task_4(C) \longleftarrow &~(~Artifact_2(C)~) \\ \nonumber 
                      &\otimes ~~~+ Artifact_5(C). \\ \nonumber 
Task_4(C) \longleftarrow &~(~Artifact_4(C)~) \\ \nonumber 
                      &\otimes ~~~+ Artifact_5(C).
\end{align}

Suppose that in a sample software development project, none of the artifacts $ Artifact_1 $, $ Artifact_2 $, $ Artifact_3 $, $ Artifact_4 $, and $ Artifact_5 $ has been produced for the component $ car $ yet. If the project manager wants to check the possibility of the execution of $ Task_4(car) $, he can simply execute the following query in the knowledge base:

\begin{equation}
?- \diamond~Task_4(car).
\end{equation}

Clearly, Transaction Logic responds $ false $ to this query as none of the preconditions of $ Task_4(car) $ has been produced so far. If an execution of a task inserts $ Artifact_1(car) $ to the knowledge base, then the response to that query will be $ true $.

\chapter{A Case Study} \label{chap:case_study}

\begin{figure}
  \centering
  \includegraphics[scale=.35]{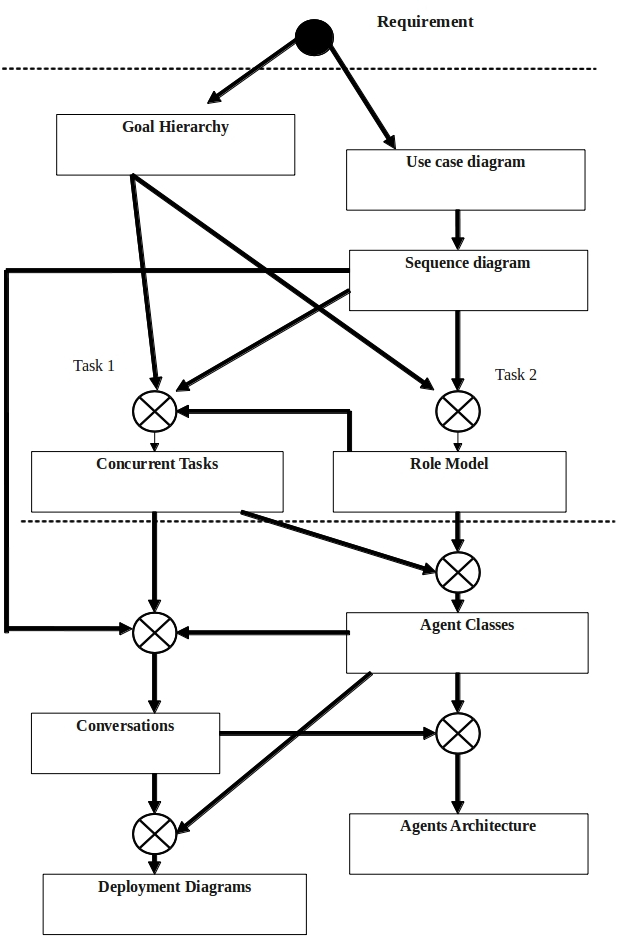}
  \caption{The life-cycle of MaSE methodology.}
  \label{fig:MaSE}
\end{figure}

In this Chapter, we briefly explain our method using a case study. We use the life-cycle of \emph{MaSE} and \emph{MASCommonKADS}, two famous agent oriented software engineering methodologies. Figure~\ref{fig:MaSE} shows the sequence of tasks in MaSE. For instance, $ Task~1 $ in Figure~\ref{fig:MaSE} can be represented as follows:

\begin{align}
Task_1 \longleftarrow &~(~Goal\_Hierarchy \wedge \\ \nonumber 
                      &~~~Sequence\_diagram \wedge \\ \nonumber  
                      &~~~ Role\_Model~) \\ \nonumber 
                      &\otimes ~~~+ Concurrent\_Tasks.
\end{align}

\begin{figure}
  \centering
  \includegraphics[scale=.35]{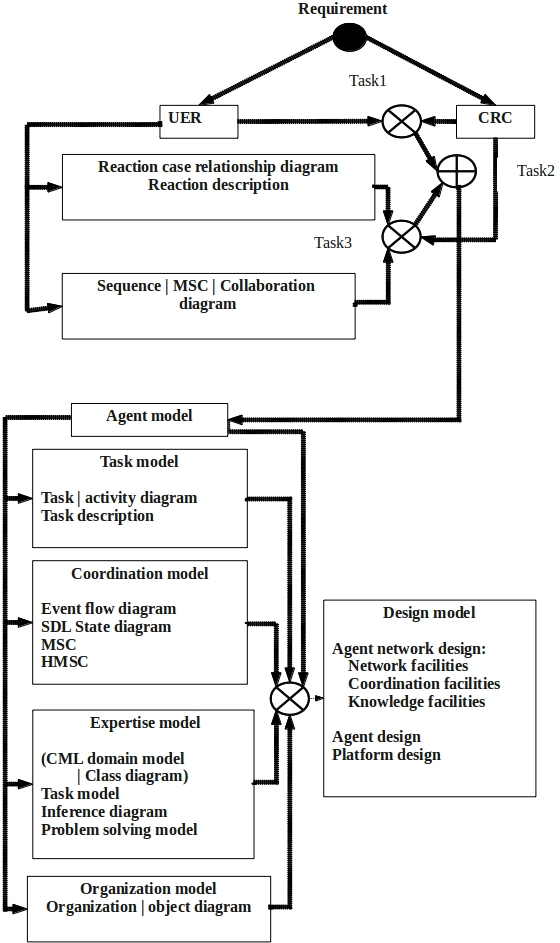}
  \caption{The life-cycle of MASCommonKADS methodology.}
  \label{fig:MASCommonKADS}
\end{figure}

Figure~\ref{fig:MASCommonKADS} also shows the sequence of tasks in MASCommonKADS. For instance, the combination of $ Task~1 $, $ Task~2 $, and $ Task~3 $  in Figure~\ref{fig:MASCommonKADS} can be defined as $ Task~4 $, represented as follows:

\begin{align}
Task_4 \longleftarrow &~(UER \wedge CRC~) \\ \nonumber 
                      &\otimes ~~~+ Agent\_Model.  \\ \nonumber
Task_4 \longleftarrow &~(~Reaction\_diagram \wedge \\ \nonumber 
                      &~~~ Collaboration\_diagram~) \\ \nonumber 
                      &\otimes ~~~+ Agent\_Model.  
\end{align}

\chapter{Conclusion} \label{chap:conclusion}
In this report, we discussed a novel method for representing software development life-cycles in Transaction Logic. As an illustration, we have shown that sophisticated
software development life-cycles, such as MaSE's life-cycle, can be naturally represented in Transaction Logic and that the use of this powerful logic opens up new possibilities for generalizations and devising new, more efficient methodologies. This representation also can be used in the development of CASE tools. 

We are planning to investigate using defeasible reasoning and argumentation theory \cite{DBLP:conf/iclp/FodorK11} to extend our developments. Another promising direction for this research is to investigate other rule based system, e.g. Answer Set Programming \cite{DBLP:conf/iclp/GelfondL88}\cite{DBLP:journals/aicom/GebserKKOSS11}, and possibly get more efficient methodologies.

\bibliographystyle{plain}
\bibliography{SoftwareWorkflowModelingReport}

\end{document}